\def\year{2022}\relax
\documentclass[letterpaper]{article}
\usepackage{aaai22}
\usepackage{times}
\usepackage{helvet}
\usepackage{courier}
\usepackage[hyphens]{url}
\usepackage{graphicx}
\usepackage{natbib}
\usepackage{caption}
\DeclareCaptionStyle{ruled}{labelfont=normalfont,labelsep=colon,strut=off}
\usepackage{amsmath,amssymb}
\usepackage{xspace}
\usepackage{xcolor}
\usepackage[labelformat=simple]{subfig}
	
\usepackage{booktabs}
\usepackage[ruled,linesnumbered]{algorithm2e}
  
  \SetCommentSty{commentstyle}
\usepackage{mdframed}
\makeatletter
\patchcmd{\@algocf@start}
  {-1.5em}
  {-.5mm}
{}{}
\makeatother

\nocopyright

\title{Verification of Neural-Network Control Systems \\ by Integrating Taylor Models and Zonotopes}
\author {
    Christian Schilling,\textsuperscript{\rm 1}
    Marcelo Forets,\textsuperscript{\rm 2}
    Sebasti\'{a}n Guadalupe\textsuperscript{\rm 2}
}
\affiliations {
    \textsuperscript{\rm 1} Aalborg University, Denmark \\
    \textsuperscript{\rm 2} Universidad de la Rep\'{u}blica, Uruguay \\
    christianms@cs.aau.dk, mforets@gmail.com, sebastianguadalupe00@gmail.com
}

\newcommand{\R}{\ensuremath{\mathbb{R}}\xspace}
\newcommand{\nat}{\ensuremath{\mathbb{N}}\xspace}
\newcommand{\plant}{\ensuremath{f}\xspace}
\newcommand{\nn}{\ensuremath{N}\xspace}
\newcommand{\pc}{\ensuremath{p2c}\xspace}
\newcommand{\cp}{\ensuremath{c2p}\xspace}
\newcommand{\period}{\ensuremath{\tau}\xspace}
\newcommand{\sol}{\ensuremath{\xi}\xspace}
\newcommand{\dom}{\ensuremath{\mathcal{D}}\xspace}
\newcommand{\rem}{\ensuremath{\Delta}\xspace}
\newcommand{\X}{\ensuremath{\mathcal{X}}\xspace}
\newcommand{\Y}{\ensuremath{\mathcal{Y}}\xspace}
\newcommand{\reach}{\ensuremath{\mathcal{S}}\xspace}
\newcommand{\hreach}{\ensuremath{\mathcal{\overline{\reach}}}\xspace}
\newcommand{\U}{\ensuremath{\mathcal{U}}\xspace}
\newcommand{\TM}{\ensuremath{\mathcal{T}}\xspace}
\newcommand{\TMR}{\ensuremath{\mathcal{R}}\xspace}
\newcommand{\Z}{\ensuremath{\mathcal{Z}}\xspace}
\newcommand{\Hy}{\ensuremath{\mathcal{H}}\xspace}

\newcommand{\ra}[1]{\ensuremath{\rho_{#1}}\xspace}
\newcommand{\rap}{\ensuremath{\ra{\plant}}\xspace}
\newcommand{\ran}{\ensuremath{\ra{\nn}}\xspace}
\newcommand{\hl}[1]{\ensuremath{\mathbf{#1}}\xspace}
\newcommand{\asgn}{\ensuremath{\leftarrow}\xspace}

\begin{document}

\maketitle

\begin{abstract}
We study the verification problem for closed-loop dynamical systems with neural-network controllers (NNCS).
This problem is commonly reduced to computing the set of reachable states.
When considering dynamical systems and neural networks in isolation, there exist precise approaches for that task based on set representations respectively called Taylor models and zonotopes.
However, the combination of these approaches to NNCS is non-trivial because, when converting between the set representations, dependency information gets lost in each control cycle and the accumulated approximation error quickly renders the result useless.
We present an algorithm to chain approaches based on Taylor models and zonotopes, yielding a precise reachability algorithm for NNCS.
Because the algorithm only acts at the interface of the isolated approaches, it is applicable to general dynamical systems and neural networks and can benefit from future advances in these areas.
Our implementation delivers state-of-the-art performance and is the first to successfully analyze all benchmark problems of an annual reachability competition for NNCS.
\end{abstract}

\section{Introduction}

In this work we consider controlled dynamical systems where the plant model is given as a nonlinear ordinary differential equation (ODE) and the controller is implemented by a neural network.
We call such systems \emph{neural-network control systems} (NNCS).
We are interested in reachability properties of NNCS: guaranteed reachability of target states or non-reachability of error states.
These questions can be verified by computing a set that overapproximates the reachable states, which is the subject of reachability analysis, with a large body of works for ODEs \cite{AlthoffFG20} and neural networks \cite{LiuALSBK21}.

In principle, reachability analysis for NNCS can be implemented by chaining two off-the-shelf tools for analyzing the ODE and the neural network.
The output set of one tool is the input set to the other, and this process is repeated for each control cycle.
This idea is indeed applied by several approaches \cite{TranYLMNXBJ20,ClaviereAGP21}.
While correct, such an approach often yields sets that are too conservative to be useful in practice.
The reason is that with each switch to the other tool, a conversion between set representations is required because the tools use different techniques.
Thus some of the dependency information encoded in the sets is lost when the tools are treated as black boxes.
This incurs an approximation error that quickly accumulates over time, also known as the wrapping effect \cite{Neumaier93}.

\smallskip

Reachability algorithms at a sweet spot between precision and performance in the literature are based on Taylor models for ODEs \cite{MakinoB03,ChenAS12} and on set propagation via abstract interpretation \cite{CousotC77} for neural networks, particularly using zonotopes \cite{GehrMDTCV18,SinghGMPV18}.
In this work we propose a new reachability algorithm for NNCS that combines Taylor models and zonotopes.
In general, Taylor models and zonotopes are incomparable and cannot be converted exactly.
We describe how to tame the approximation error when converting between these two set representations with two main insights.
First, we identify a special structured zonotope, which can be exactly converted to a Taylor model by encoding the additional structure in the so-called remainder.
Second, the structure of the Taylor model from the previous cycle can be retained by only updating the control inputs, which allows to preserve the dependencies encoded in the Taylor model.

Our approach only acts at the set interface and does not require access to the internals of the reachability tools.
They only need to expose the complete set information, which is only a minor modification of the black-box algorithms.
Thus our approach makes no assumptions about the ODE or the neural network, as long as there are sound algorithms for their (almost black-box) reachability analysis available.
This makes the approach a universal tool.
While our approach is conceptually simple, we demonstrate in our evaluation that it is effective and scalable in practice.
We successfully analyzed all benchmark problems from an annual NNCS competition \cite{ARCHCOMP} for the first time.

\smallskip

In summary, this paper makes the following contributions:

\begin{itemize}
	\item We propose structured zonotopes and show how to soundly convert them to Taylor models and back.
	
	\item We design a sound reachability algorithm for NNCS based on Taylor models and zonotopes.
	
	\item We demonstrate the precision and scalability of the algorithm on benchmarks from a reachability competition.
\end{itemize}

\subsection{Related Work}

The verification of continuous-time NNCS has recently received attention.
The tool \emph{Verisig} \cite{IvanovWAPL19} transforms a neural network with sigmoid activation functions into a hybrid automaton \cite{AlurCHH92} and then uses \emph{Flow$^*$} \cite{ChenAS13}, a reachability tool for nonlinear ODEs based on Taylor models, to analyze the transformed control system as a chain of hybrid automata.
While that approach allows to preserve dependencies in the Taylor model, it is not applicable to the common ReLU activation functions and the automaton's dimension scales with the number of neurons.
The tool \emph{NNV} \cite{TranYLMNXBJ20} combines \emph{CORA} \cite{Althoff15}, a reachability tool for nonlinear ODEs based on (variants of) zonotopes, and an algorithm based on star sets \cite{TranBXJ20} for propagating through a ReLU neural network.
That approach suffers from the loss of dependencies when switching between the set representations.
The tool \emph{Sherlock} \cite{DuttaCS19,DuttaCJST19} combines \emph{Flow$^*$} with an output-range analysis for ReLU neural networks \cite{DuttaJST18a}.
That approach abstracts the neural network by a polynomial, which has the advantage that dependencies can in principle be preserved.
While the approach requires hyperrectangular input sets and the abstraction comes with its own error, this approach can be precise in practice for small input sets.
The input to the neural network as well as the polynomial order must be small in practice for scalability reasons.
Further, the approach only works well for neural networks with a single output; for multiple outputs, the analysis has to be applied iteratively.
The tool \emph{ReachNN$^*$} \cite{HuangFLC019,FanHCL020} approximates Lipschitz-continuous neural networks with Bernstein polynomials and then analyzes the resulting polynomial system with \emph{Flow$^*$}; estimates of the Lipschitz constant tend to be conservative.
\citet{ClaviereAGP21} combine validated simulations and abstract interpretation.

A number of approaches consider discrete-time systems, which are considerably easier to handle.
\citet{XiangTRJ18} study the simple case of discrete-time piecewise-linear (PWL) systems and controllers with ReLU activation functions, for which one can represent the exact reachable states as a union of convex polytopes.
However, the number of polytopes may grow exponentially in the dimension of the neural network.
\emph{VenMAS} \cite{AkintundeBKL20} also assumes PWL dynamics and considers a multi-agent setting with a temporal-logic specification.
\citet{DuttaJST18b} consider nonlinear dynamics and compute a template polyhedron that overapproximates the output of the neural network based on range analysis.
\emph{OVERT} \cite{SidraneMIK21} approximates nonlinear dynamics by PWL bounds.
\citet{BacciG021} consider unbounded time.

\paragraph{Outline}

In the next section we continue with the background on NNCS and set representations. Afterward we describe our approach and evaluate it, and finally we conclude.


\section{Preliminaries}

We formally introduce NNCS and the core set representations used in our approach: Taylor models and zonotopes.

\subsection{Neural-Network Control Systems}

We consider plants modeled by ODEs $\dot{x} = \plant(x, u)$ where $x \in \R^n$ is the state vector and $u \in \R^m$ is the vector of control inputs.
Given an initial state $x(0) = x_0$ and a (constant) control input $u_0$, we assume that the solution of the corresponding initial-value problem at time $t \geq 0$, denoted by $\sol(t, x_0, u_0, \plant)$, exists and is unique.
(We only discuss deterministic plants here to simplify the presentation.
The extension to nondeterministic systems is straightforward; handling such systems is common in the reachability literature \cite{SingerB06,AlthoffFG20} and orthogonal to the problem described in this work.)

A neural-network control system (NNCS) is a tuple $(\plant, \nn, \pc, \cp, \period)$ with a plant $\plant(x, u)$ over $x \in \R^n$ and $u \in \R^m$, a controller given as a neural network $\nn : \R^i \to \R^o$, a function $\pc : \R^n \to \R^i$ that takes the current plant state $x$ and turns it into the input to the controller, a function $\cp : \R^o \to \R^m$ that takes the controller output and turns it into the new control input $u$, and a control period $\period \in \R_{> 0}$.

\begin{figure}[tb]
	\centering
	\includegraphics{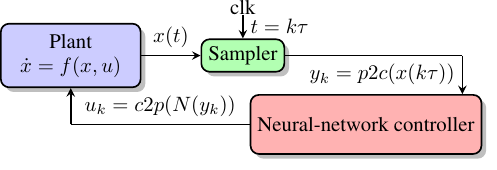}
	\caption{Neural-network control system.}
	\label{fig:nncs}
\end{figure}

A conceptual sketch of an NNCS is given in Figure~\ref{fig:nncs}.
The NNCS periodically queries the controller for new control inputs.
At time points $k \period$, $k \in \nat$, the state $x(k \period)$ is passed to $\pc$, to the controller $\nn$, and to $\cp$, which yields the new control inputs $u_k$.
Here we use the common assumption that the computation of $u_k$ is instantaneous.
The sequence of the $u_k$ induces a continuous piecewise input signal $u(t)$.
Formally, given an initial state $x_0$ at $t = 0$, we recursively define the sequence of input vectors $u_k$, $k \in \nat$, and the evolution of the state $x(t)$, $t \geq 0$, which is a trajectory of the NNCS:
\begin{align}
	u_k &= \cp(\nn(\pc(x(k \period)))) \label{eq:uk} \\
	x(t) &=
	\begin{cases}
		x_0 & t = 0 \\
		\sol(t - k \period, x(k \period), u_k, \plant) & t \in (k \period, (k+1) \period]
	\end{cases}
	\label{eq:xt}
\end{align}

We may also write $x(t, x_0)$ resp.\ $u_k(x_0)$ for the trajectory $x(t)$ resp.\ for the vector $u_k$ to clarify the dependency on $x_0$.

\begin{figure}
	\centering
	\subfloat[Zonotope.]{
		\centering
		\includegraphics[width=.48\linewidth,keepaspectratio]{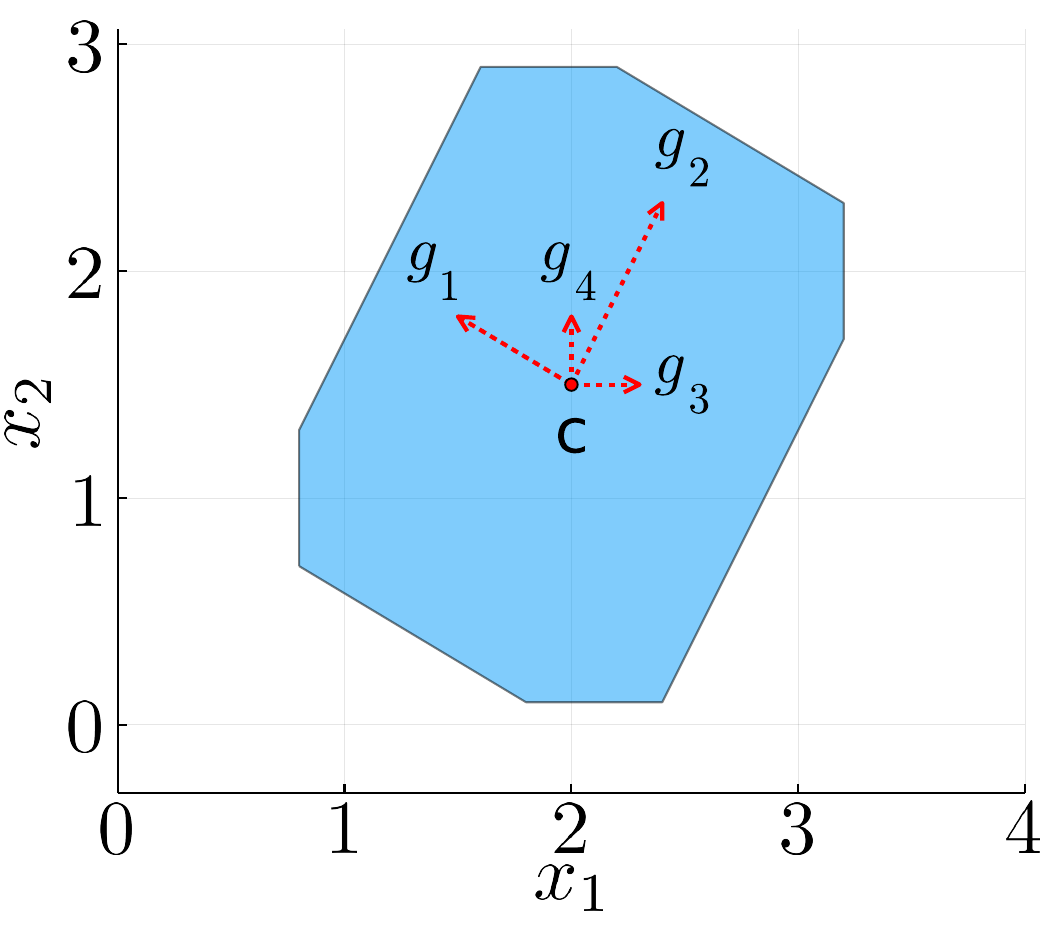}
		\label{fig:zonotope}
	}%
	\subfloat[Taylor model.]{
		\centering
		\includegraphics[width=.48\linewidth,keepaspectratio]{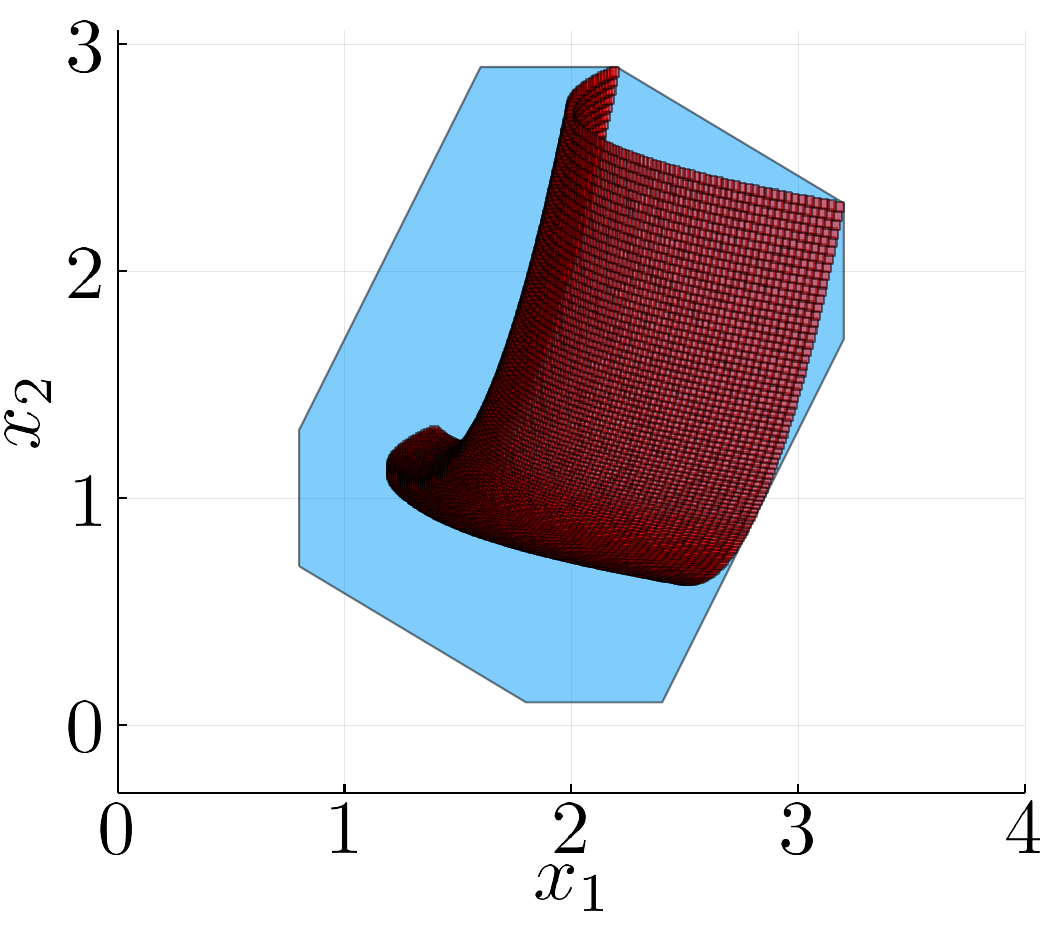}
		\label{fig:TM_zonotope_conversion}
	}
	\caption{A structured zonotope (left) and a Taylor model covered with one zonotope and with a union of boxes (right).}
	\label{fig:TM_zonotope}
\end{figure}

\subsection{Taylor Models}

A $d$-dimensional Taylor model of order $k$ is a tuple $\TM = (p, \rem, \dom)$ where $p = (p_1, \dots, p_d)^T$ is a vector of multivariate polynomials $p_i : \dom \to \R$ of degree at most $k$, $i = 1, \dots, d$, the remainder $\rem = \rem_1 \times \dots \times \rem_d$ is a hyperrectangle containing the origin, and $\dom \subseteq \R^d$ is the domain \cite{MakinoB03}.
Thus $\TM$ represents the vector-valued function $p(x) + \rem$: an interval tube around the polynomial $p$.
We often use the common normalization $\dom = [-1, 1]^d$, which can be established algorithmically.

\paragraph{Example}

The tuple $(p, \Delta, \dom)$ with $p_1(x) = -0.5 x^2 + 3 x$ over domain $\dom = [-1, 1]$ and remainder $\rem = [-0.1, 0.1]$ is a one-dimensional Taylor model of order $2$.

\subsection{Taylor-Model Reach Sets (TMRS)}

A Taylor-model reach set (TMRS) $\TMR$ is a structure used in reachability algorithms when propagating Taylor models through an ODE in time. 
For a $d$-dimensional system, a TMRS is a $d$-vector of Taylor models in one variable representing time with shared domain.
The coefficients of these Taylor models are themselves multivariate polynomials in the $d$ state variables, whose domain is assumed to be the symmetric box $[-1, 1]^d$.
(The time domain of a TMRS is not normalized to $[-1, 1]$.)
Evaluating a TMRS over a time point (or a time interval) yields a $d$-dimensional Taylor model.

\paragraph{Example}

Continuing the previous example, consider the one-dimensional TMRS consisting of the Taylor model $(q, \Delta, [0, 1])$ where $q_1(t) = (- 0.5 x^2 + 2 x) t + x$.
Evaluation at $t = 1$ yields the Taylor model from the previous example.

\subsection{Zonotopes}

A zonotope is the image of a hypercube under an affine transformation and hence a convex centrally-symmetric polytope \cite{Ziegler95}.
Zonotopes are usually characterized in generator representation:
An $n$-dimensional zonotope $\Z \subseteq \R^n$ with center $c \in \R^n$ and $p$ generators $g_j \in \R^n$ ($j = 1, \dots, p$) is defined as
\[
	\Z = \left\{ x \in \mathbb{R}^n : x = c + \sum_{j=1}^p \zeta_j g_j ~,~ \zeta_j \in [-1, 1] \right\}.
\]
The generators are commonly aligned as columns in a matrix $G_\Z = \begin{bmatrix} g_1 & g_2 & \cdots & g_p \end{bmatrix}$.
The order of $\Z$ is the ratio $p / n$ of generators per dimension.
In the special case that $G_\Z$ is a diagonal matrix, the zonotope represents a hyperrectangle.

Given two sets $\X_1, \X_2 \subseteq \R^n$, their Minkowski sum is
\[
	\X_1 \oplus \X_2 = \{ x_1 + x_2 : x_1 \in \X_1, x_2 \in \X_2 \}
\]
We say that a zonotope $\Z$ with center $c$ and generator matrix $G_\Z$ is \emph{structured} if it has order $2$ and $G_\Z$ has the block structure $\begin{bmatrix} M & D \end{bmatrix}$, where $D$ is a diagonal matrix.
A structured zonotope corresponds to the Minkowski sum of 1)~the zonotope centered in $c$ with generator matrix $M$ and 2)~the hyperrectangle centered in the origin whose radius corresponds to the diagonal of $D$.

\paragraph{Example}

The two-dimensional structured zonotope with the center $c = (2, 1.5)^T$ and the generator matrix $\begin{pmatrix} -0.5 & 0.4 & 0.3 & 0 \\ 0.3 & 0.8 & 0 & 0.3 \end{pmatrix}$ is depicted in Figure~\ref{fig:zonotope}.

\paragraph{Propagating zonotopes through a neural network}

Abstract interpretation \cite{CousotC77} is a well-known technique to propagate sets through a system in a sound (i.e., overapproximate) way.
The sets are taken from a class called the abstract domain.
The idea is to compute the image of the set under the system's successor function; if the image does not fall into the abstract domain, an overapproximation from that domain is chosen.
For neural networks the idea is to iteratively propagate a set through each layer.
For instance, several algorithms based on the zonotope abstract domain \citep{GhorbalGP09} for propagation through neural networks have been proposed \cite{GehrMDTCV18,SinghGMPV18}.
Given an input zonotope, the algorithm outputs a zonotope that overapproximates the exact image of the neural network.
The smallest zonotope overapproximation is not unique.
Zonotopes are efficient to manipulate and closed under the affine map in each layer (multiplication with the weights and addition of the bias); only the activation function requires an overapproximation.


\section{Reachability Algorithm}

In this section we formalize the reachability problem for NNCS, explain how one can convert between (structured) zonotopes and Taylor models in a sound way, and finally integrate these conversions into a reachability algorithm.

\subsection{Problem Statement}

Given an NNCS, a set of initial states $\X_0 \subseteq \R^n$, and another set of states $\Y \subseteq \R^n$, we are interested in answering two types of questions:
The \emph{must-not-reach} question asks whether no trajectory reaches any state in $\Y$.
The \emph{must-reach} question asks whether each trajectory reaches some state in $\Y$.
To answer these questions, we aim at computing the reachable states
$
	\reach_t = \{x(t, x_0) : x_0 \in \X_0\}
$
at time $t$.
Determining reachability of a state (i.e., membership in $\reach_t$) is undecidable, which follows from undecidability of reachability for nonlinear dynamical systems \cite{Hainry08}.

The common approach in the literature is to consider the reachable states for time intervals $[T_0, T_1]$,
$
	\reach_{[T_0, T_1]} = \bigcup_{t \in [T_0, T_1]} \reach_t,
$
and to find a coverage $\hreach_{[T_0, T_1]} \supseteq \reach_{[T_0, T_1]}$.
This allows to give one-sided guarantees: if $\hreach_{[0, T]} \cap \Y = \emptyset$ for some time horizon $T$, we can affirmatively answer a \emph{must-not-reach} question, and if $\hreach_{[T_0, T_1]} \subseteq \Y$ for some time interval $[T_0, T_1]$, we can affirmatively answer a \emph{must-reach} question.
The task we aim to solve is thus, given a time horizon $T \geq 0$, to compute a covering $\hreach_{[0, T]} \supseteq \reach_{[0, T]}$.

\subsection{Computing Sequences of Covering Sets}

Given a set of initial states $\X_0 \subseteq \R^n$ and a bound on the number of control periods $K \in \nat$, for each $k = 0, \dots, K$ we want to cover the set of control inputs $\U_k = \{u_k(x_0) : x_0 \in \X_0\}$ by a set $\Z_k \supseteq \U_k$.
Similarly, we want to compute sets $\TMR_k \supseteq \reach_{[k \period, (k+1) \period]}$.
The idea is to represent the sets $\Z_k$ as structured zonotopes and the sets $\TMR_k$ as TMRS.
From the recursive definition in \eqref{eq:uk} and \eqref{eq:xt}, for each control cycle we shall evaluate the TMRS, obtaining a Taylor model, then convert to a zonotope and back, and finally compute a new TMRS.
Exact conversion between Taylor models and zonotopes is generally not possible, so we must overapproximate.

To simplify the presentation, we turn the input variables into new state variables by adding $m$ fresh state variables with zero dynamics.
Thus from now on we sometimes assume a state vector of dimension $n + m$.

\subsection{From Taylor Model to Structured Zonotope}

Given an $n$-dimensional Taylor model $\TM$, we want to compute a covering zonotope.
(Since a Taylor model can represent nonlinear dependencies and a zonotope only consists of linear constraints, one cannot hope for an exact conversion.)
We construct a structured zonotope $\Z$ as the Minkowski sum $\Z_l \oplus \Hy_\mathit{nl}$ of a zonotope and a hyperrectangle.
The intuition is that $\Z_l$ exactly captures the linear part of the polynomial and $\Hy_\mathit{nl}$ overapproximates the nonlinear part and the remainder.

We split the Taylor model's polynomial vector into the linear part $p_l = A x + b$ (for some $A \in \R^{n \times n}, b \in \R^n$) and the nonlinear part $p_\mathit{nl}$.
The zonotope $\Z_l$ has the center $b$ corresponding to the constant term of $p_l$ and the generator matrix $A$ corresponding to the linear coefficients.
Recall that the domain of $p_l$ is normalized to $[-1, 1]^n$, so it conforms with the zonotope's definition.
Then we compute the interval approximation $\Hy'$ of $p_\mathit{nl}$ by evaluating the polynomials over the domain using interval arithmetic.
Finally we define $\Hy_\mathit{nl} = \Hy' \oplus \rem$, where $\rem$ is the remainder of $\TM$.

\paragraph{Example}

Let the Taylor model $\TM = (p, [0, 0]^2, [-1, 1]^2)$ with polynomials $p_1(x) = 0.6 x_1^2 - 0.5 x_1 + 0.4 x_2 + 1.7$ and $p_2(x) = 0.6 x_2^2 + 0.3 x_1 + 0.8 x_2 + 1.2$.
We obtain the zonotope $\Z$ from the previous example.
Figure~\ref{fig:TM_zonotope_conversion} shows $\Z$ together with a multi-box cover of $\TM$, for which we split the domain into $10{,}000$ uniform boxes and evaluate $\TM$ using interval arithmetic.
Note that $\Z$ is tight at multiple edges.

\subsection{From Structured Zonotope to Taylor Model}

Reachability algorithms for nonlinear dynamical systems based on Taylor models assume that the set of initial states $\X_0$ itself is given as a Taylor model.
Converting hyperrectangles to a Taylor model is easy.
Hence the typical approach is to first overapproximate $\X_0$ with a hyperrectangle.
However, this way we lose all dependencies between variables.
If we were to apply this conversion in each control cycle, the approximation error would quickly explode.
We describe a better approximation when $\X_0$ is a zonotope.
Zonotopes of order $> 1$ generally cannot be converted exactly to a Taylor model (see \cite[Corollary~1]{KochdumperA21} applied to zonotopes).
Here we show that for structured zonotopes (which have order $2$) an exact conversion is possible.

Say we are given a structured zonotope $\Z \subseteq \R^m$ (the new control inputs).
We construct the Taylor model $\TM = (p, \rem, [-1, 1]^m)$ corresponding to $\Z$, for which it remains to describe how to construct each $p_j$ and $\rem_j$, $j = 1, \dots, m$.

First we explain the construction in the simpler case that $\Z$ is a hyperrectangle $\Hy$ with center $c$ and radius $r$.
Then each $p_j$ is the constant polynomial $c_j$, where $c_j$ is the $j$-th component of $c$, with the remainder $\rem_j = [-r_j, r_j]$ (i.e., $\rem$ is the hyperrectangle $\Hy$ with the center shifted to the origin).

Now we explain the construction in the case that $\Z$ is a structured zonotope with generator matrix $G_\Z = \begin{bmatrix} M & D \end{bmatrix}$.
Let $A_{i, j}$ denote the entry of matrix $A$ at row $i$ and column $j$.
We define the polynomial $p_j(x) = c_j + \sum_{k=1}^{m} M_{j, k}\,x_k$ (which is correct because we use the domain $[-1, 1]^m$) and the remainder $\rem_j = [-d, d]$ where $d = |D_{j, j}|$.

Observe that this conversion is exact and compatible with the other conversion algorithm.
Let $\Z$ be a structured zonotope.
Then converting to a Taylor model and back using our algorithms yields $\Z$ again.

\paragraph{Example}

We continue with the structured zonotope $\Z$ from the previous example. The corresponding Taylor model is $p_1(x) = 2.0 - 0.5 x_1 + 0.4 x_2$ and $p_2(x) = 1.5 + 0.3 x_1 + 0.8 x_2 + [-0.3, 0.3]$, with $\rem_1 = \rem_2 = [-0.3, 0.3]$.

\subsection{Reachability Algorithm Based on Taylor Models and Zonotopes}

Now we have all ingredients to formulate a reachability algorithm.
Initially we are given an NNCS $(\plant, \nn, \pc, \cp, \period)$ as defined before, a set of initial states $\X_0 \subseteq \R^n$, and a time horizon $T \in \R_{> 0}$.
If $\X_0$ is not already given as a Taylor model, we need to convert (e.g., from a structured zonotope) or overapproximate it first.
We assume two black-box reachability algorithms $\ran$ and $\rap$.
Algorithm $\ran$ receives a zonotope $\Z$ and produces another zonotope that covers the image of $\Z$ under the controller $\nn$, e.g., implementing the algorithm in \cite{SinghGMPV18}.
Algorithm $\rap$ receives a Taylor model and a time horizon and produces a TMRS covering the reachable states of the plant $\plant$, e.g., implementing the algorithm in \cite{MakinoB03}.

\begin{algorithm}[tb]
	\caption{Reachability algorithm for NNCS.}
	\label{alg:reach}
	\KwIn{%
	\begin{minipage}[t]{67mm}
		$(\plant, \nn, \pc, \cp, \period)$: NNCS;
		$\X_0$: initial states;
		$T = K \period$: time horizon;
		$\ran$: reachability algorithm for $\nn$;
		$\rap$: reachability algorithm for $\plant$
	\end{minipage}}
	\KwOut{TMRS overapproximating the reachable states until $T$}
	\medskip
	$\TM_x \asgn \textit{TaylorModel}(\X_0)$\tcp*{construct Taylor model from $\X_0$}
	
	\For{$k \asgn 0$ \KwTo $K-1$}{%
		$\Z' \asgn \pc(\textbf{to$\_$Zonotope}(\TM_x))$%
		\tcp*{convert to zonotope}
		\label{line:step1-3}
		
		$\Z \asgn \cp(\ran(\Z'))$%
		\tcp*{zonotope covering $\nn$'s output}
		\label{line:step2}
		
		$\TM_u \asgn \textbf{to$\_$TM}(\Z)$%
		\tcp*{convert to Taylor model}
		\label{line:step3}
		
		$\TM' \asgn \textit{merge}(\TM_x, \TM_u)$\tcp*{$n+m$-dimensional Taylor model by merging Taylor models} \label{line:step4-1}
		
		$\TMR_k \asgn \rap(\TM', \period)$\tcp*{TMRS covering reachable states for one control cycle} \label{line:step4-2}
		
		$\TM \asgn \textit{evaluate}(\TMR_k, k \period)$\tcp*{Taylor model at next sampling time point} \label{line:step1-1}
		
		$\TM_x \asgn \textit{project}(\TM, [1, \dots, n])$\tcp*{project Taylor model to the state variables} \label{line:step1-2}
	}
	\Return{$(\TMR_0, \dots, \TMR_{K-1})$}
\end{algorithm}

Algorithm~\ref{alg:reach} consists of a loop of four main steps.
We assume that the time horizon $T$ is a multiple $K$ of the period $\period$.
(For other time horizons one can just execute the loop body once more with a shortened time frame in line~\ref{line:step4-2}.)
Say we are in iteration $k$.
The first step is to obtain the control inputs for the next time period from the controller.
According to \eqref{eq:uk}, we need to extract the current state information, which is stored as part of a Taylor model $\TM$.
We obtain $\TM$ by evaluating the current TMRS at $t = k \period$ (lines~\ref{line:step1-1} and \ref{line:step1-2}).
Then we convert $\TM$ to a structured zonotope $\Z'$ using our algorithm described above (line~\ref{line:step1-3}).
Finally we apply the function $\pc$ and pass the set to the second step.
In typical cases such as affine maps ($\pc(x) = A x + b$), we can apply $\pc$ directly to $\Z'$.
In more complicated cases we could instead apply $\pc$ to the Taylor model before the conversion.

The second step is to propagate $\Z'$ through the controller via $\ran$ (line~\ref{line:step2}).
The output is a new zonotope $\Z$.
Then we apply the function $\cp$ to it; again this is easy for affine maps, and otherwise we need to overapproximate.

The third step is to construct a Taylor model $\TM_u$ from $\Z$, using our algorithm from above (line~\ref{line:step3}), and then merge with $\TM_x$, the first $n$ dimensions of $\TM$, to obtain an $n+m$-dimensional Taylor model.
This works because $\TM_x$ does not depend on the inputs.
To obtain a structured zonotope $\Z$, we use the order-reduction algorithm in \cite{Girard05}.

The fourth step is to propagate the TMRS through the plant via $\rap$ for the next $\period$ time frame (lines~\ref{line:step4-1} and~\ref{line:step4-2}).


\section{Evaluation}

We implemented the algorithm in JuliaReach, a toolbox for reachability analysis \cite{BogomolovFFPS19}.
Set representation and set conversion is implemented in the library LazySets \cite{LazySets}.
For the Taylor-model analysis we use the implementation by \citet{TaylorSeriesJOSS,BenetFSS19}.
For the zonotope propagation we implemented the algorithm by \citet{SinghGMPV18}.
To obtain simulations in the visualizations we use the ODE solver by \citet{RackauckasN19}.
All results reported here were obtained on a standard laptop with a quad-core 2.2\,GHz CPU and 8\,GB RAM running Linux.

We consider the benchmark problems used in the competition on NNCS at ARCH-COMP 2021 \cite{ARCHCOMP}.
In total there are seven problems with various features.
The problems have up to $12$ continuous states, $5$ hidden layers, and $500$ hidden units.
All problems use ReLU activation functions.
We exclude one problem from the presentation because it differs in scope (linear discrete-time behavior and multiple controllers).
Next we study one of the problems in detail.
Then we report on the results for the other problems.
The experiments are available at \textit{https://github.com/JuliaReach/AAAI22\_RE/}.

\subsection{Case Study: Unicycle Model}

\begin{figure}[tb!]
	\centering
	\includegraphics[width=\linewidth,keepaspectratio]{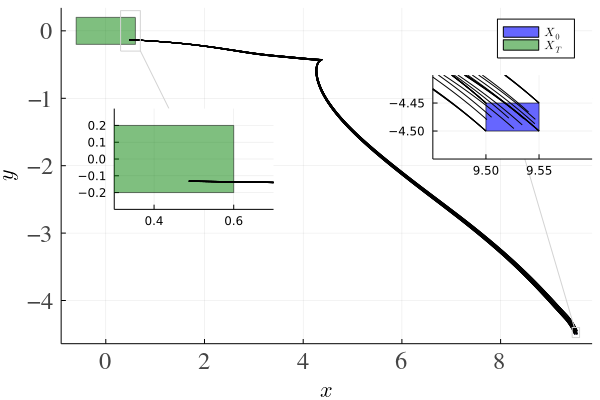}
	
	\includegraphics[width=\linewidth,keepaspectratio]{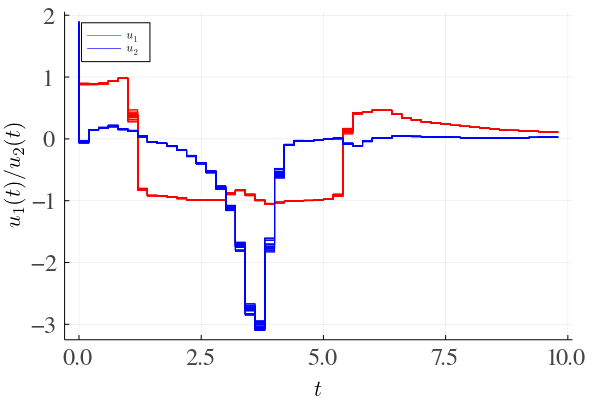}
	\caption{Simulations for the unicycle model: projection in state space (first plot) and control signals (second plot).}
	\label{fig:unicycle_sim}
\end{figure}

We consider the model of a unicycle, which was originally used in \cite{DuttaCS19}.
The plant has four state variables $(x, y, \theta, v)$, where $x$ and $y$ represent the wheel coordinates in the plane, $\theta$ is the yaw angle of the wheel, and $v$ is the velocity.
There are two inputs $(u_1, u_2)$, where $u_1$ controls the acceleration and $u_2$ controls the wheel direction.
Finally, there is a disturbance $w$.
The dynamics are given as the following system of ODEs:
\begin{align*}
	\dot{x} &= v \cos(\theta) &
	\dot{y} &= v \sin(\theta) &
	\dot{\theta} &= u_2 &
	\dot{v} &= u_1 + w
\end{align*}

A neural-network controller with one hidden layer ($500$ neurons) was trained with a model-predictive control scheme as teacher.
The function $\pc$ is the identity, while the controller output is post-processed with $(u_1, u_2)^T = \cp((o_1, o_2)^T) = (o_1 - 20, o_2 - 20)^T$.
The controller is sampled with a period $\period = 0.2\,s$.
The uncertain set of initial states $\X_0$ is given by $x \in [9.5, 9.55]$, $y \in [-4.5, -4.45]$, $\theta \in [2.1, 2.11]$, $v \in [1.5, 1.51]$, and $w \in [-10^{-4}, 10^{-4}]$.
The specification is to reach a target set $\X_T$ given by $x \in [-0.6, 0.6]$, $y \in [-0.2, 0.2]$, $\theta \in [-0.06, 0.06]$, $v \in [-0.3, 0.3]$ within a time horizon of $T = 10\,s$.

In Figure~\ref{fig:unicycle_sim} we show $\X_0$, $\X_T$, and ten random simulations together with simulations from all $32$ extremal points of $\X_0$ and the domain of $w$.
We can see that $\X_T$ is reached only in the last moment, so the analysis requires a high precision to prove containment of the reachable states at $t = 10\,s$.
Our implementation can prove containment for three state variables, but the lower bound for $y$ slightly exceeds $-0.2$.

We found that the zonotope approximation is suboptimal for this controller.
To improve precision, we reduce the dependency uncertainty in the initial states by splitting $\X_0$ into $3 \times 1 \times 8 \times 1 = 24$ smaller hyperrectangles.
Then we have to solve $24$ reachability problems, where the final reachable states are the union of the individual results.
Mathematically, these sets are equivalent, but set-based analysis generally gains precision from smaller initial states.
We note that the analysis is embarrassingly parallelizable, but our current implementation does not make use of that.

\begin{figure}[tb]
	\centering
	\includegraphics[width=\linewidth,keepaspectratio]{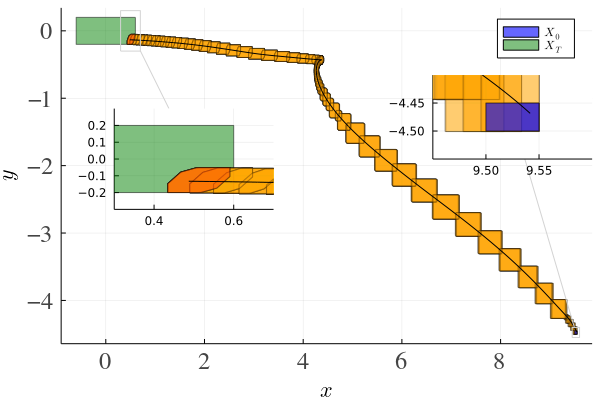}
	\caption{Reachability analysis for the unicycle model.}
	\label{fig:unicycle_reach}
\end{figure}

Using an adaptive step size with absolute tolerance $10^{-15}$ and order-$10$ Taylor models, we can verify the property within $93$ seconds (i.e., four seconds for each sub-problem).
The reach sets of all $24$ runs together with a random simulation are shown in Figure~\ref{fig:unicycle_reach}.
The reach sets for different sub-problems overlay each other soon after the beginning, indicating that the controller quickly steers trajectories from different sources to roughly the same states.
We could have evaluated the final TMRS at the time point $t = 10$ for higher precision, but this was not required.

\smallskip

\begin{figure}[tb]
	\centering
	\includegraphics[width=.49\linewidth,keepaspectratio]{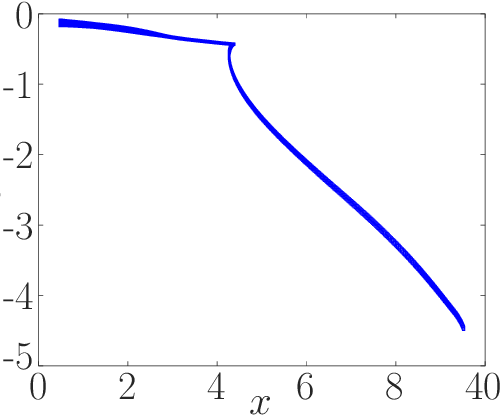}
	\includegraphics[width=.49\linewidth,keepaspectratio]{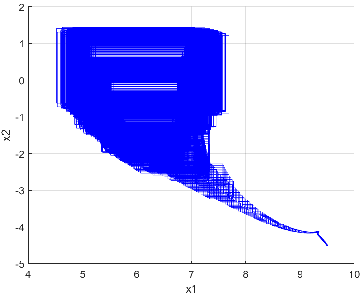}
	\caption{Results from Sherlock (left) and NNV (right; only partial result) on the unicycle model ($x$/$y$ projection).}
	\label{fig:unicycle_other_tools}
\end{figure}

We compare to the results of \emph{Sherlock} \cite{DuttaCS19}, which originally proposed the benchmark problem.
(We are not aware of any tool that can handle all benchmark problems.)
As discussed in the related work, that approach can be very precise because it does not have to switch between set representations, and indeed there is no splitting required here.
However, controllers with multiple outputs need to be handled by repeating the analysis for each output neuron individually, which is costly.
In total the analysis takes $525$ seconds and produces the plot in Figure~\ref{fig:unicycle_other_tools}.

We also compare to the result of NNV \cite{TranYLMNXBJ20}, which uses a black-box reachability method for the plant and hence loses many dependencies.
In \cite{ARCHCOMP} the authors reported that their tool runs out of memory before the analysis finishes.
The plot in Figure~\ref{fig:unicycle_other_tools} contains intermediate results, which show that the precision declines quickly.

\subsection{Other Problems From ARCH-COMP}

\begin{table}[tb]
	\centering
	\begin{tabular}{@{\,} l @{} c @{\ } c @{\ \ } r @{\ \ } r @{\,}}
		\toprule
		Problem & Dimensions & Cyc. & Sherlock & JuliaReach \\
		\midrule
		Unicycle & $\hl{4};500;\hl{2}$ & $50$ & $526$ & $93$ \\
		TORA & $\hl{4};100,100,100;\hl{1}$ & $20$ & $30$ & $2040$ \\
		ACC & $\hl{6};20,20,20,20,20;\hl{1}$ & $50$ & $4$ & $1$ \\
		S.\ pend. & $\hl{2};25,25;\hl{1}$ & $20$ & $1$ & $1$ \\
		D.\ pend. & $\hl{4};25,25;\hl{2}$ & $20$ & $6^\dag$ & $4$ \\
		Airplane & $\hl{12};100,100,20;\hl{6}$ & $20$ & $169^\dag$ & $29$ \\
		\bottomrule
	\end{tabular}
	\caption{Benchmarks.
	The second column shows the number of state variables $n$, neurons per hidden layer, and control variables $m$.
	The other columns show the number of control cycles and the the run time of Sherlock resp.\ JuliaReach in seconds (averaged over five runs, rounded to integers).
	A ``$\dag$'' marks measurement until the tool stopped working.}
	\label{tab:benchmarks}
\end{table}

\begin{figure}[tb!]
	\centering
	\subfloat[TORA.]{
		\centering
		\includegraphics[width=.49\linewidth,height=3cm]{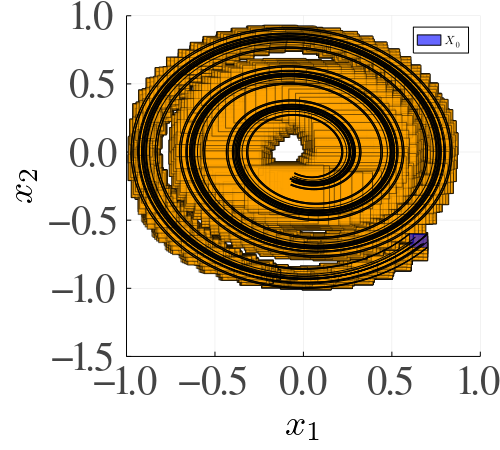}
		~
		\includegraphics[width=.45\linewidth,height=3cm]{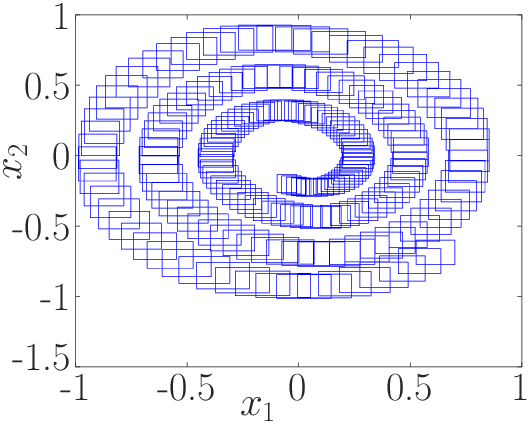}
		\label{fig:tora}
	}
	
	\subfloat[ACC.]{
		\centering
		\includegraphics[width=.49\linewidth,height=3cm]{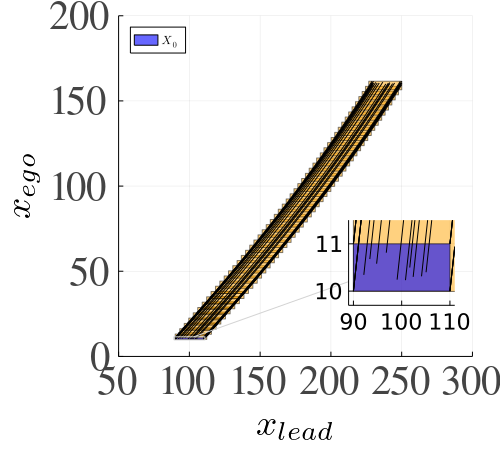}
		~
		\includegraphics[width=.45\linewidth,height=3cm]{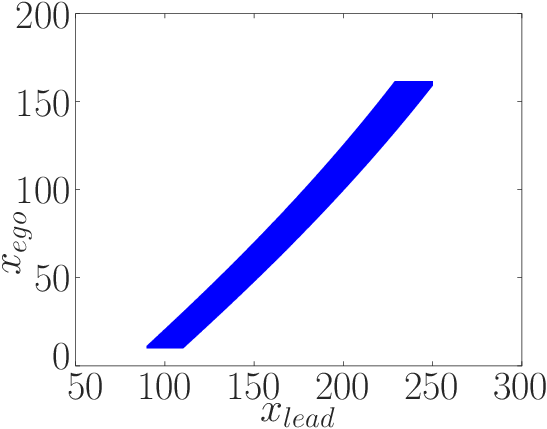}
		\label{fig:acc}
	}
	
	\subfloat[Single pendulum.]{
		\centering
		\includegraphics[width=.49\linewidth,height=3cm]{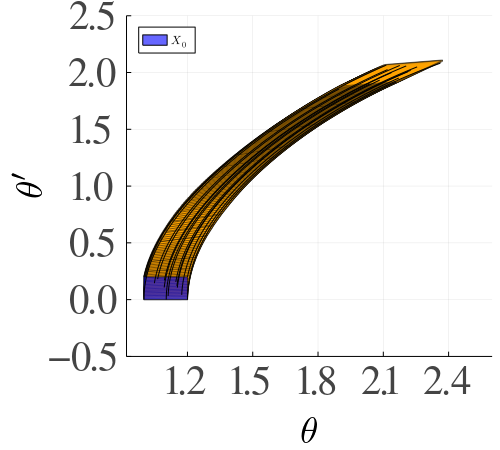}
		~
		\includegraphics[width=.45\linewidth,height=3cm]{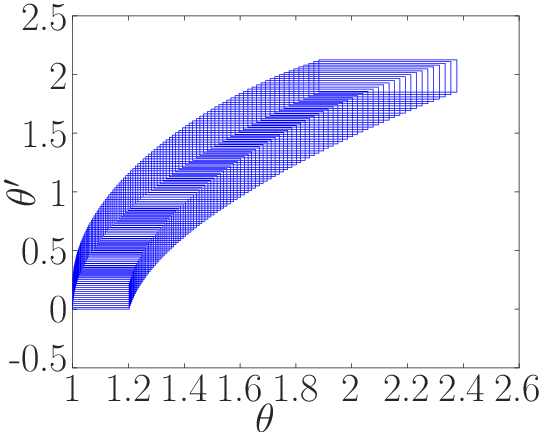}
		\label{fig:singlependulum}
	}
	
	\subfloat[Double pendulum. Left: Additional result from a smaller initial set (in red). Right: Incomplete result from the smaller initial set.]{
		\includegraphics[width=.49\linewidth,height=3cm]{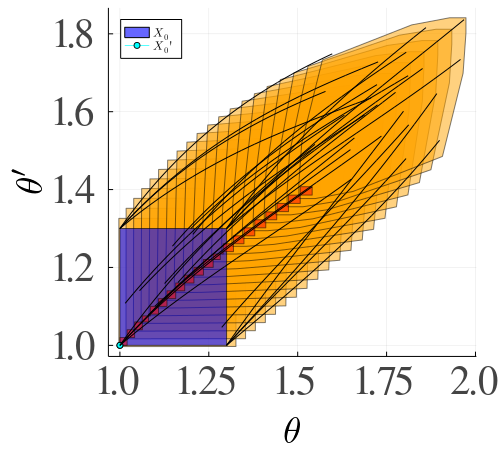}
		~
		\includegraphics[width=.45\linewidth,height=3cm]{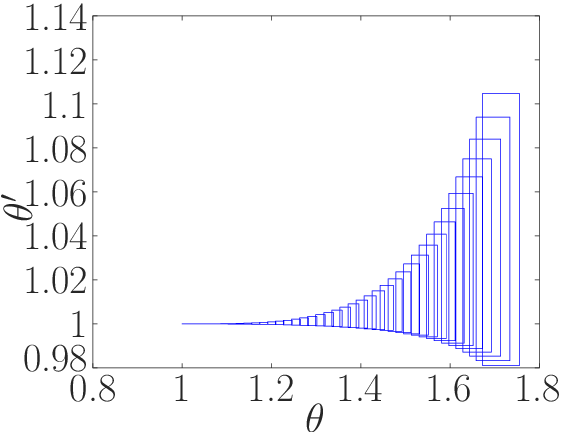}
		\label{fig:doublependulum}
	}
	
	\subfloat[Airplane. Right: Incomplete result.]{
		\centering
		\includegraphics[width=.49\linewidth,height=3cm]{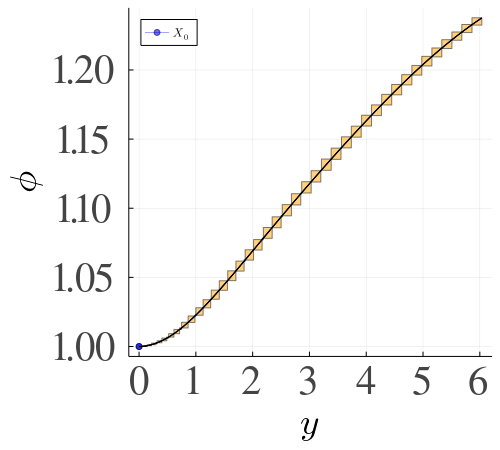}
		~
		\includegraphics[width=.45\linewidth,height=3cm]{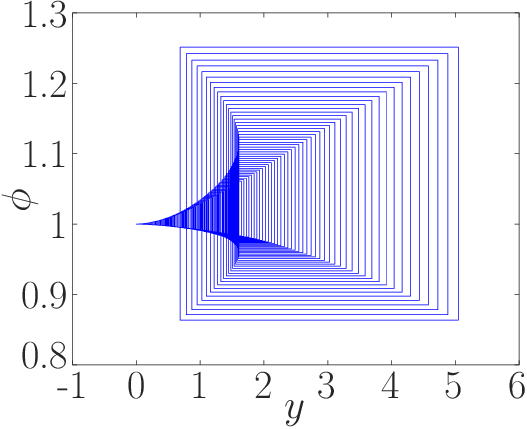}
		\label{fig:airplane}
	}
	\caption{Results from JuliaReach (left) and Sherlock (right).
	We additionally plot simulations from the extreme points of $\X_0$ and ten additional random points.}
	\label{fig:comparison}
\end{figure}

Here we shortly summarize the results on the remaining benchmark problems from ARCH-COMP 2021: a translational oscillator with a rotational actuator (TORA), an adaptive cruise control (ACC), a single and a double pendulum, and an airplane model.
We summarize the core model properties in Table~\ref{tab:benchmarks}.
Since NNV cannot solve many of the problems, we only discuss the results of our implementation and Sherlock.
The reachability results are plotted in Figure~\ref{fig:comparison}.

The TORA problem was also proposed by the Sherlock authors.
Here $\cp(u) = u - 10$.
Again the zonotope algorithm produces relatively coarse results and our tool JuliaReach has to split heavily to yield the required precision, which makes it slow.
For all other problems, JuliaReach is precise enough without splitting and is faster than Sherlock.

The implementation of Sherlock requires ReLU activation functions at every layer, including the output layer.
This is not common, but we modified the controllers of the last four problems accordingly for a fair comparison.
The original specifications are not satisfied by the modified controllers and hence we only compare reachability results and run time here for these problems.
Our implementation can solve all benchmark problems with the original controllers, as shown in the ARCH-COMP 2021 report \cite{ARCHCOMP}.

For the ACC problem, $\pc(x) = (30, 1.4, x_5, x_1 - x_4, \linebreak[1] x_2 - x_5)^T$ (an affine map); the tools have similar precision but ours is faster.
For the single-pendulum problem, our implementation is more precise.
On the remaining two problems with multiple control variables (double pendulum and airplane), Sherlock diverges.
As discussed before, Sherlock uses an approximate analysis in those cases.
For the double-pendulum problem, it stops after the first control cycle and returns with an error about divergence.
Splitting $\X_0$ helps for a while, but Sherlock diverges after seven control cycles even when splitting each dimension into $30{,}000$ pieces.
For the airplane problem, the trajectories obtained with the modified controller expand fast and hence we define a smaller initial set.
Sherlock is slow due to the large number of control variables and diverges after ten control cycles.

\subsection{Discussion}

To summarize, our approach generally produces precise results, as can be seen from the simulations in the plots covering most of the reachable states, and can additionally benefit from splitting the initial set; for Sherlock, the effect of splitting is smaller and does not help solving the double pendulum and airplane problems.
Sherlock is typically precise and sufficiently fast for controllers with a single output, although not always more precise than our implementation.
For multiple outputs, Sherlock is slower and often diverges.

Our algorithm relies on two reachability algorithms and inherits their scalability.
We shortly discuss the most relevant parameters for NNCS reachability.
1)~Plant dimension: Reachability methods for high-dimensional nonlinear systems are generally not available.
2)~ Neural-network dimension: The algorithm from \cite{SinghGMPV18} scales to realistic neural networks in control applications.
3)~Number of iterations: Each iteration incurs a conversion between set representations, which makes the task more challenging.


\section{Conclusion}

In this paper we have addressed the reachability problem for neural-network control systems.
When combining successful reachability tools for the ODE and neural-network components, the main obstacle is the conversion of sets at the tool interface.
We have proposed a conversion scheme when the ODE analyzer uses Taylor models and the neural-network analyzer uses zonotopes.
Our approach is able to preserve most dependencies between the control cycles.
Our implementation is the first to successfully analyze all benchmark problems of the verification competition ARCH-COMP 2021.
Compared to Sherlock, our approach works reliably for neural networks with multiple output dimensions.

For future work, we plan to investigate the interface for other set representations.
For example, CORA \cite{Althoff15} can use polynomial zonotopes \cite{Althoff13,KochdumperA21}, which are as expressive as Taylor models and conversion to and from zonotopes works well.
Another direction is to combine different reachability algorithms, e.g., the one in \cite{GehrMDTCV18} or the ``polynomialization'' approach used in Sherlock \cite{DuttaCS19}; thus we could compute several output sets and then choose one or even combine them.


\section*{Acknowledgments}

We thank Luis Benet for fruitful discussions about the Taylor-model implementation.
We also thank the anonymous reviewers at AAAI 2022 for their feedback.
S.G.\ acknowledges funding from Julia Seasons of Contributions.

\bibliography{bibliography}

\end{document}